\documentclass[aip,jcp,reprint,noshowkeys]{revtex4-1}
\usepackage{graphicx,tabularx,dcolumn,bm,xcolor,microtype,hyperref,multirow,amsmath,wrapfig,amssymb,amsfonts,physics,float,soul,rotating,longtable}
\usepackage[version=4]{mhchem}

\newcolumntype{d}{D{.}{.}{-1}}
\usepackage[normalem]{ulem}

\definecolor{darkgreen}{RGB}{0, 180, 0}

\newcommand{\mcc}[1]{\multicolumn{1}{c}{#1}}

\newcommand{\Eadia}{E^\text{adia}}
\newcommand{\EOO}{E^\text{0-0}}
\newcommand{\Ea}{E_\text{abs}^\text{vert}}
\newcommand{\Ef}{E_\text{fluo}^\text{vert}}
\newcommand{\EreorgES}{E_\text{reorg}^\text{ES}}
\newcommand{\EreorgGS}{E_\text{reorg}^\text{GS}}

\newcommand{\Ezpve}{\Delta E^\text{ZPVE}}

\newcommand{\ChemAcc}{\%{\text{CA}}}
\newcommand{\AccErr}{\%{\text{AE}}}

\newcommand{\LCPQ}{Laboratoire de Chimie et Physique Quantiques (UMR 5626), Universit\'e de Toulouse, CNRS, UPS, France}
\newcommand{\CEISAM}{Laboratoire CEISAM (UMR 6230), CNRS, Universit\'e de Nantes, Nantes, France}

\begin{document}

\title{Chemically Accurate 0-0 Energies with not-so-Accurate Excited State Geometries}

\author{Pierre-Fran\c{c}ois Loos}
	\affiliation{\LCPQ}
\author{Denis Jacquemin}
	\email{Denis.Jacquemin@univ-nantes.fr}
	\affiliation{\CEISAM}

\begin{abstract}
Using a series of increasingly refined wavefunction methods able to tackle electronic excited states, namely ADC(2), CC2, CCSD, CCSDR(3) and CC3, we investigate
the interplay between geometries and 0-0 energies. We show that, due to a strong and nearly systematic error cancelation between the vertical 
transition and geometrical reorganization energies, CC2 and CCSD structures can be used to obtain chemically-accurate 0-0 energies, though the underlying geometries are
rather far from the reference ones and would deliver significant errors for many chemical and physical properties. This indicates that obtaining 0-0 energies matching 
experiment does not demonstrate the quality of the geometrical parameters.  In contrast, accurate computation of vertical excitation energies is mandatory in order to reach chemical accuracy.
By determining CC3 total energies on CCSD structures, we model a large set of compounds (including radicals) and electronic transitions (including singlet-triplet excitations) and 
successfully reach chemical accuracy in a near systematic way. Indeed, for this particular set, our protocol delivers a mean absolute error as small as $0.032$ eV,
chemical accuracy (error smaller than $1$ kcal.mol$^{-1}$ or $0.043$ eV) being obtained in 80\%\ of the cases.
In only three cases the error exceeds $0.15$ eV which is of the order of the typical error provided by TD-DFT or second-order wavefunction methods for this particular property. 
The present composite protocol is therefore very effective despite the fact that the geometries may not be considered as very accurate.
\begin{figure}[H]
	\centering 
	\includegraphics[width=0.5\linewidth]{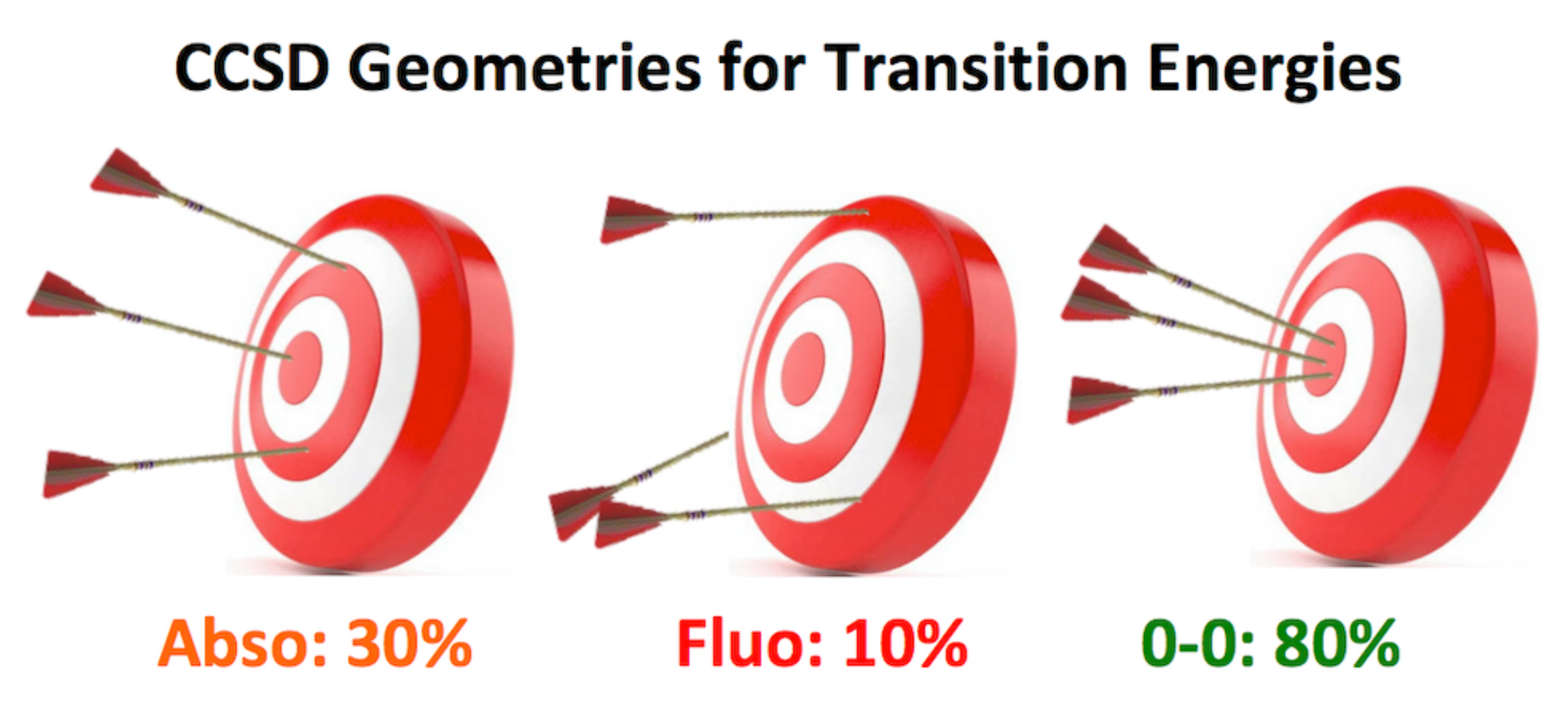}
\end{figure}
\end{abstract}

\maketitle

% 
% I. Introduction
%
\section{Introduction}
\begin{figure}[htp]
  \includegraphics[width=0.5\textwidth]{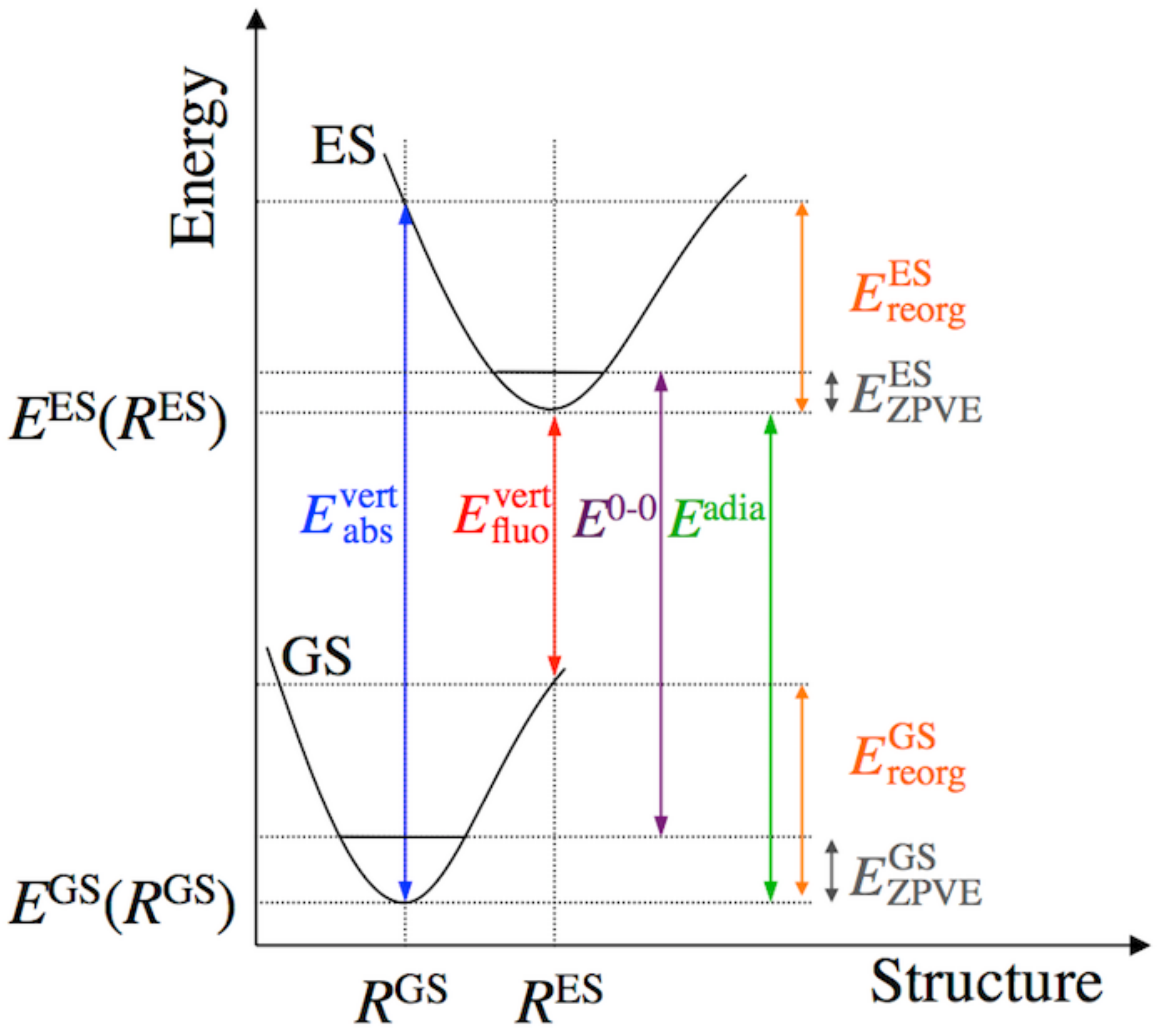}
  \caption{Representation of transition energies and energy differences discussed in the present work. $\Ea$ (blue) and $\Ef$ (red) are the (vertical) absorption and emission/fluorescence energies, while
  $\EreorgGS$ and $\EreorgES$ (orange) are the (geometrical) reorganization energies of the GS and ES states, respectively.
  The adiabatic and 0-0 energies are represented in green and purple, respectively. 
  All these energies are defined as positive quantities.
 }
   \label{Fig-1}
\end{figure}

The accurate modeling of phenomena occurring in electronic excited states (ESs) is often required to attain an in-depth understanding of experimental observations made in, e.g., solar cells and light emitting diodes. 
However, for such materials, a straightforward relationship between measured and computed ES properties is often hampered by the complexity of the chemical system (surrounding environment, non-adiabatic processes, 
dynamical effects, \ldots).  This limits both the quality of the theory that can be applied and the experimental result precision. Therefore, when one is interested in direct theory-experiment comparisons, the most valuable target property 
probably remains the 0-0 energy ($\EOO$) because $\EOO$ (often denoted $T_{00}$ or ``band origin'' experimentally): i) has been measured for many molecules in gas phase with uncertainty typically smaller than 1 cm$^{-1}$, and 
ii) is a well-defined theoretical quantity which corresponds to the difference between the ES and ground-state (GS) energies taken at their respective geometrical minimum (the adiabatic energy, $\Eadia$), corrected 
by their corresponding zero-point vibrational energies (ZPVEs) (see Figure \ref{Fig-1}).  This contrasts with many other ES properties, such as experimental bond distances and dipole moments that are often obtained indirectly and therefore come 
with significant error bars, or vertical transition energies which can be easily computed but have no clear experimental counterpart.

Given that $\EOO$ offers a meaningful comparison between theory and experiment, it is not surprising that many benchmarks have been devoted to their modeling. 
\cite{Fur02,Die04b,Gri04b,Rhe07,Hel08,Goe10a,Sen11b,Jac12d,Win13,Jac14a,Fan14b,Jac15b,Oru16,Loo18b} These benchmarks have been performed considering either gas-phase molecules, for which theory--experiment comparisons are 
straightforward, \cite{Fur02,Die04b,Gri04b,Rhe07,Hel08,Sen11b,Win13,Fan14b,Oru16,Loo18b} or solvated dyes, for which the measured absorption-fluorescence crossing point (AFCP) can be taken as a reference. 
\cite{Die04b,Goe10a,Jac12d,Jac14a,Jac15b} The second approach allows tackling larger compounds, of higher interest for practical applications, but at the cost of additional challenges originating, as mentioned above, from the modeling of the surrounding environment.  Irrespectively 
of the set of compounds, the vast majority of these benchmarks have been carried out with time-dependent density functional theory (TD-DFT), \cite{Ulr12b} or correlated wavefunction approaches (partially) including contributions from
double excitations, i.e., the configuration interaction singles with perturbative doubles [CIS(D)],\cite{Hea94} the second-order algebraic diagrammatic construction [ADC(2)],\cite{Dre15} and the second-order coupled-cluster-based CC2 method.\cite{Chr95}  
Although the corresponding results are unsurprisingly dependent on both the theoretical protocol and the selected set of molecules, the overall accuracy of the final theoretical estimates, as measured by the mean absolute error (MAE) with respect
to experiment, typically falls in the $0.10$--$0.30$ eV window, i.e., far from the desired ``chemical accuracy'' ($1.0$ kcal.mol$^{-1}$  or $0.043$ eV error). Very recently, we have proposed a protocol reaching, for organic compounds, such accuracy on an almost systematic
basis. \cite{Loo18b} Given that the vibrational correction is known to be relatively insensitive to the selected method, \cite{Jac12d,Win13,Bud17} this protocol mainly focusses on the accurate calculation of adiabatic energies, 
$\Eadia$. To this end, high-level coupled-cluster methods including contributions from the triples, i.e., CCSDR(3)/\emph{def2}-TZVPP and CC3/\emph{aug}-cc-pVTZ, have been respectively applied to obtain the geometrical parameters 
and the vertical transition energies. This led to a MAE of $0.018$ eV for a set of 35 singlet-singlet valence states of small organic molecules, well below the chemical accuracy threshold.\cite{Loo18b} However, in the same work, we also showed  that
computing the transition energies at a high-level of theory, i.e., all-electron CC3 calculations in our case, is required to achieve a small MAE. Indeed, keeping the same geometries but determining $\Eadia$ at the CCSDR(3), CCSD, 
and CC2 levels led to MAEs of  $0.046$, $0.207$, and $0.078$ eV, respectively, whereas freezing the core electrons during the CC3 calculations was enough to double the MAE to $0.045$ eV. \cite{Loo18b}

In the present contribution, we do assess the impact of the geometries on theoretical $\EOO$ values. This question arises because the calculation of CCSDR(3)/\emph{def2}-TZVPP geometries was the clear computational bottleneck of our original protocol. \cite{Loo18b}
Indeed, this method not only includes perturbative corrections for the triples, as CCSD(T) for the GS, which comes with a non-favorable scaling with system size, but, in addition, does not have analytic gradients implemented which
means that the gradient minimization process had to be carried out purely numerically. It would be undoubtedly much more advantageous to be able to use CCSD, CC2, or ADC(2) structures, as this would both decrease the scaling with system size  
and also allow taking advantage of analytical gradients.  Whilst the potential benefit was clear,  hope was dim!  To understand why, let us first consider two exact formulations of the adiabatic energy: \cite{Jac12d}
\begin{subequations}
\begin{align}
	\Eadia & = E^\text{ES} (R^\text{ES}) -  E^\text{GS} (R^\text{GS}), 	
	\label{eq:1} \\
	           & = \frac{\Ea + \Ef}{2} +  \frac{\EreorgGS - \EreorgES}{2}. 	
	\label{eq:2}
\end{align}
\end{subequations}
The first equation gives the standard minimum-to-minium energy difference definition of $\Eadia$. However, as one can see in Eq.~\eqref{eq:2}, $\Eadia$ can be also expressed as the average of the absorption and fluorescence 
vertical energies corrected by half of the \emph{difference} between the GS and ES geometrical reorganization energies (see also Figure \ref{Fig-1}).  While this second definition does not offer a more efficient expression for practical calculations, 
it helps analyzing methodological trends. Indeed, except for compounds exhibiting important differences in ground and excited state potential energy surfaces, one can expect that the first contribution in Eq.~\eqref{eq:2} largely dominates, so that
\begin{equation}
	\label{eq:3}
	\Eadia \simeq \frac{\Ea + \Ef}{2}
\end{equation}
is a reasonable approximation. \cite{Jac12d} Second, let us consider the results of a recent work, \cite{Jac18a}  in which we compared the vertical absorption and fluorescence energies obtained on a series of increasingly accurate 
geometries. For 24 compounds, we found that selecting second-order M{\o}ller-Plesset (MP2) [the GS equivalent of ADC(2)], CC2, and CCSD geometries in lieu of CCSDR(3) structures would yield average deviations of 
$-0.01$, $-0.06$, and $+0.05$ eV for $\Ea$, respectively. For $\Ef$, the corresponding deviations are significantly larger: $-0.03$, $-0.08$ and $+0.15$ eV with ADC(2), CC2, and CCSD ES geometries, respectively.
This illustrates that the ES structures are very sensitive to the selected electronic structure method.\cite{Bud17}  As can be seen, the errors obtained with CCSD geometries largely exceed the chemical accuracy threshold for fluorescence.  This left us rather 
circumspect before starting the present study.  Indeed, one would need these errors to be almost cancelled out by the reorganization energy difference, a much smaller term, in order to reach chemically-accurate $\Eadia$ values
with CCSD structures. Of course, this can happen if the two potential energy surfaces represented in Figure \ref{Fig-1} are shifted strictly parallel to the horizontal axis: this would strongly modify the vertical energies without altering the 
adiabatic energies. However, we have also shown that:\cite{Bud17,Jac18a} i) CCSD [CC2 and ADC(2)] has a tendency to provide too localized [delocalized] ES geometries; ii) these methods yield much larger errors in the ES than in the GS, with, e.g.,  
CCSD [CC2] mean errors of $-0.021$ and $-0.007$ [$0.030$ and $0.009$] \AA\ for the  ES and GS \ce{C=O} bond lengths, respectively; and iii) the accuracy of the various methods significantly depends on the 
nature of the bonding.  This hints that the quality of the geometries might significantly influence the quality of the corresponding $\EOO$ values.
 
In the present work, we aim at multiple goals: i) investigating the impact of geometries on the computed $\EOO$ values; ii) determining, whether or not, the protocol of Ref.~\citenum{Loo18b} can be lighten; iii) estimating if the CC3
geometries would yield significant improvements over their CCSDR(3) counterparts; and iv) extending the previous benchmark set, notably by considering radical species as well as singlet-triplet excitations.

% 
% II. Computational Details
%
\section{Computational Details}
\label{sec-met}

We have used a variety of programs to determine the optimal GS and ES geometries as well as the transition energies. In every cases, all the electrons are correlated, i.e., the frozen-core
approximation was never applied.  The ADC(2) and CC2 optimizations have been performed with the Turbomole package, \cite{Turbomole} selecting the \emph{def2}-TZVPP atomic basis set and 
applying the resolution-of-identity approximation. During these calculations, the self-consistent field Hartree-Fock (HF), second-order and geometry optimization thresholds were all tightened compared to default values by selecting values of $10^{-9}$, $10^{-7}$, and $10^{-5}$ a.u., respectively. 
ADC(2) and CC2 numerical frequency calculations were systematically performed with the  same atomic basis set. The CCSD optimizations and frequency 
calculations have been performed with Gaussian16 \cite{Gaussian16} and Psi4 \cite{Psi4} using the same \emph{def2}-TZVPP basis set. The geometry convergence threshold was systematically tightened, with 
a requested residual mean force smaller than  $10^{-5}$ a.u., whereas the CCSD (EOM-CCSD) energy convergence threshold was set to, at least, $10^{-8}$ ($10^{-7}$) a.u.~in Gaussian16 in order to obtain 
accurate analytical gradients and, consequently, accurate numerical frequencies.   The CCSDR(3) and CC3 optimizations were performed with the Dalton package \cite{dalton} using the same
basis set as for the other CC models.  These optimizations used the default convergence thresholds of Dalton. We underline that analytical gradients are not available for these two levels of theory, 
so that the  CCSDR(3) and CC3  minimizations were based on the numerical differentiation of the total energies. Several geometries used here can be found in earlier contributions. \cite{Bud17,Jac18a,Bre18a,Loo18a,Loo18b}
 Unless otherwise stated, all total and transition energies reported herein have been determined at the CC3/\emph{aug}-cc-pVTZ level (no frozen core) with the
Dalton \cite{dalton} and Psi4 \cite{Psi4} packages using default algorithms and parameters. CC3 is the \emph{de facto} gold standard for ES calculations and has recently been shown
to deliver very small errors with respect to full CI estimates for small compounds. \cite{Gar17b,Loo18a,Sce18b,Gar18} Finally, the B3LYP calculations performed to obtain (TD-)DFT ZPVE were achieved with Gaussian16,\cite{Gaussian16} 
using the \emph{ultrafine} quadrature grid.

% 
% III. Results & Discussion
%
\section{Results and Discussion}
\label{sec-res}

\subsection{Formaldehyde: a representative example}

\begin{figure*}
  \includegraphics[width=\textwidth]{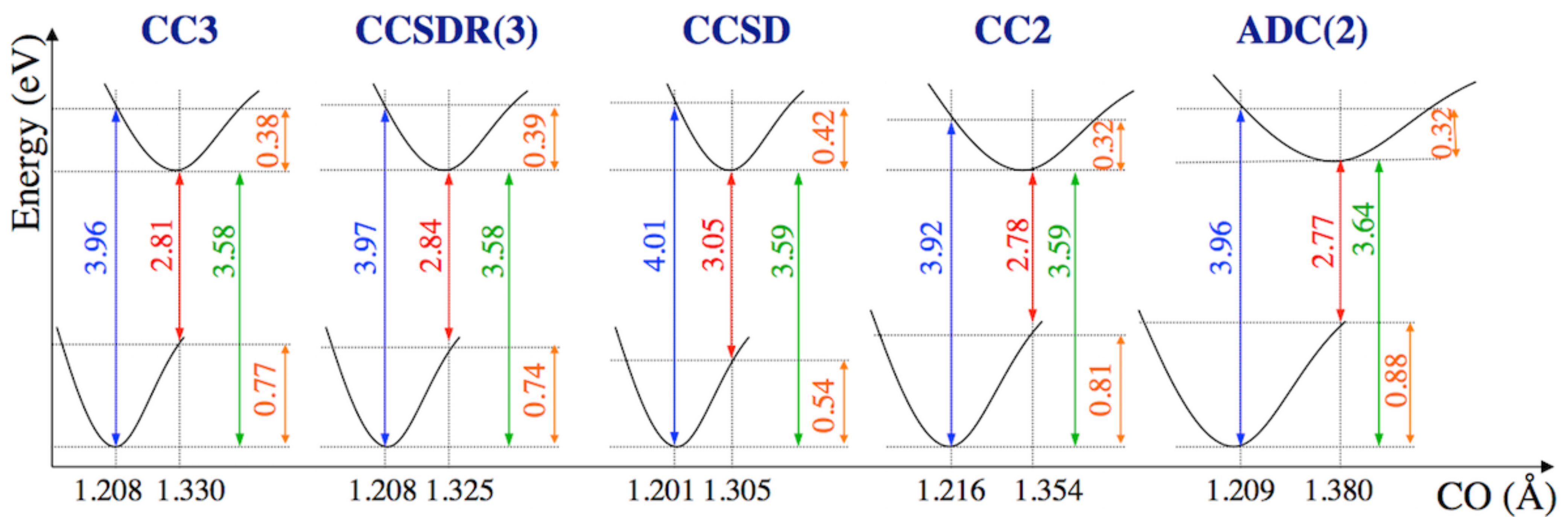}
  \caption{Representation of the CC3/\emph{aug}-cc-pVTZ transition energies of formaldehyde computed with, from left to right, the CC3, CCSDR(3), CCSD, CC2 and ADC(2) optimized geometries. 
  The absorption, fluorescence, adiabatic and reorganization energies are represented in blue, red, green and orange, respectively.
  On the horizontal axis, we provide the optimal \ce{C=O} bond lengths for these five geometries.}
  \label{Fig-2}
\end{figure*}

For illustrative purposes, we qualitatively represent, in Figure \ref{Fig-2}, CC3 potential energy surfaces obtained using five different geometries for formaldehyde, a molecule undergoing significant structural changes after
its hallmark $n \rightarrow \pi^\star$ excitation. Although such a one-dimensional representation does not provide the overall picture (e.g., the puckering angle in the ES differs significantly from one method to another), \cite{Bud17} it 
allows to qualitatively capture, for a given method, the main energetic and geometrical effects, taking the CC3 geometry as a reference.  For this geometry, $\Eadia = 3.580$ eV, a value that is, as expected (see Introduction), rather close from the 
average between vertical absorption and emission energies (3.385 eV). In formaldehyde, the planar GS is significantly stiffer than the puckered ES, and one logically finds that $\EreorgGS$ (0.77 eV) $>$ 
$\EreorgES$ (0.38 eV).

Let us now turn towards less accurate geometries. As the CCSDR(3) structures are very similar to the CC3 ones, \cite{Bud17} the CCSDR(3) and CC3 transition energies are very similar to each other as well (Figure \ref{Fig-2}), which is  
consistent with the fact that CCSDR(3) geometries were found good enough to deliver chemically accurate $\EOO$. \cite{Loo18b}  Using CCSD --- a method known to underestimate the GS-to-ES geometrical changes --- the \ce{C=O} 
bond length in the ES becomes significantly shorter. Consequently, the two vertical transition energies increase, the Stokes shift ($\Delta^{\mathrm{SS}} = \Ea - \Ef$) decreases, and the GS and ES reorganization energies become nearly 
equal. However, as one can see, $\Eadia$ remains almost unchanged. The opposite scenario is found with CC2: the elongation of the \ce{C=O} bond after excitation is exaggerated inducing an underestimation of the vertical transition 
energies, but vastly different GS and ES reorganization energies, with an overall negligible impact on $\Eadia$. Finally, considering ADC(2) which delivers a poor ES geometry for formaldehyde, the difference between $\EreorgGS$ and 
$\EreorgES$ is even more pronounced but the adiabatic energy still only moderately deviates from the CC3 reference value. In short, there is a clear error compensation mechanism between the two terms in the rhs of Eq.~\eqref{eq:2}. 
Indeed, their magnitudes significantly differ from one geometry to another with values of $3.385$, $3.405$, $3.533$, $3.350$, and $3.364$ eV for the first term, and $0.195$, $0.175$, $0.057$, $0.244$, and $0.278$ eV for the latter 
when using CC3, CCSDR(3), CCSD, CC2, and ADC(2) geometries, respectively.  Nevertheless, their sum, given by $\Eadia$, is remarkably stable: $3.580$, $3.580$, $3.589$, $3.594$ and $3.642$ eV for CC3, CCSDR(3), CCSD, 
CC2, and ADC(2), respectively.

As discussed in details below, this compensation phenomenon is rather general. The adiabatic energies are significantly less sensitive to the quality of the selected geometry than the vertical absorption and fluorescence energies. On the 
bright side, this means that one can indeed use a cheaper method than CCSDR(3) in order to get ES structures, yet reaching accurate 0-0 energies. On the dark side, this indicates that matching experiment for $\EOO$ is not a proof 
that the underlying GS and ES structures are accurate.

\subsection{Error compensation pattern}

\begin{figure*}
  \includegraphics[width=0.9\textwidth]{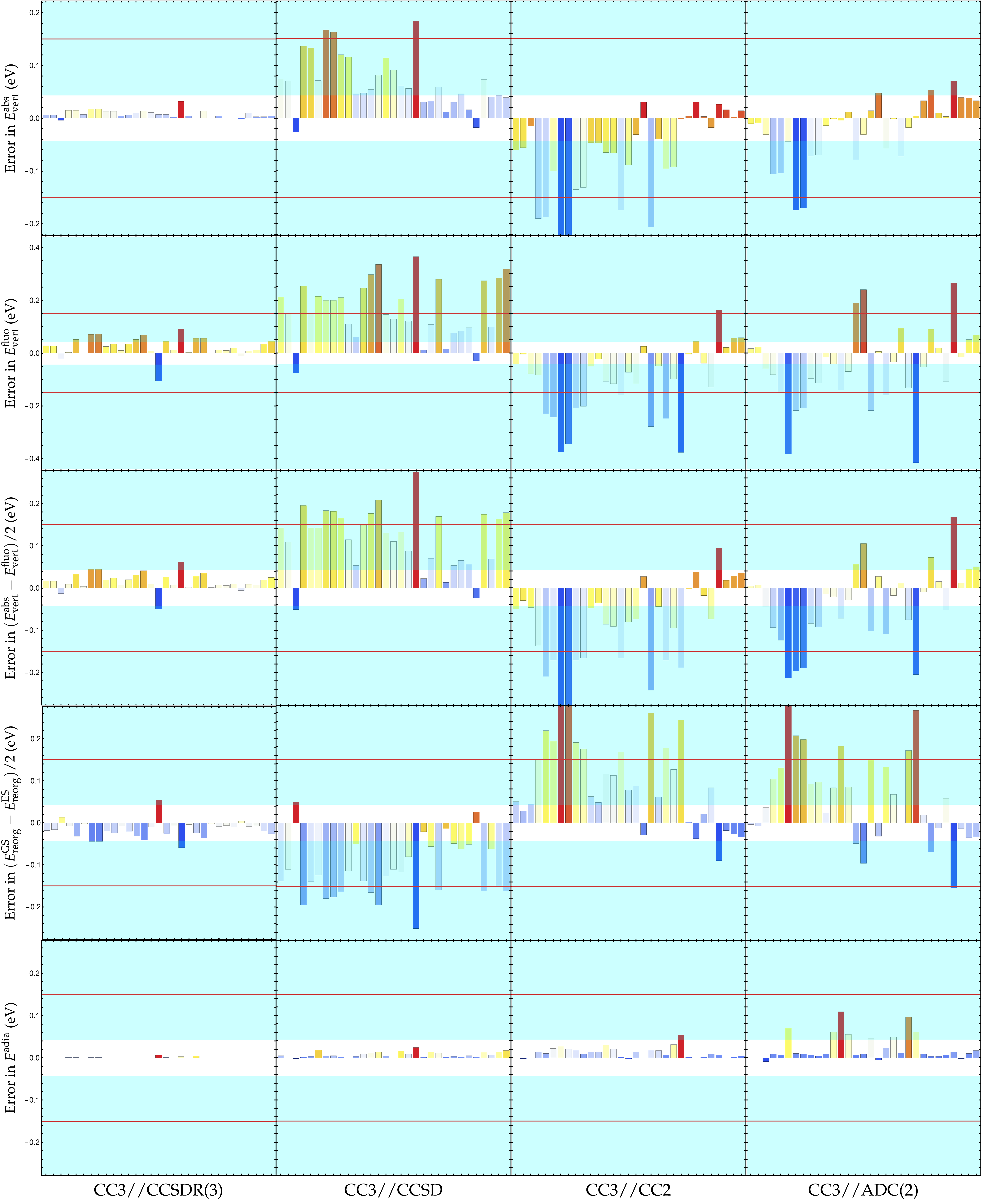}
  \caption{Bar charts of the error (in eV) made with various geometries for 31 singlet-singlet transitions using CC3/\emph{aug}-cc-pVTZ//CC3/\emph{def2}-TZVPP values as references.
	From left to right, the CCSDR(3), CCSD, CC2 and ADC(2) methods are selected for the geometry optimizations while retaining CC3 for the computation of vertical transition energies. 
	From top to bottom, $\Ea$, $\Ef$, $(\Ea + \Ef)/2$, $(\EreorgGS - \EreorgES)/2$, and $\Eadia$.   
	The white regions indicate chemical accuracy (with respect to the CC3 reference values), the horizontal red lines indicate errors of $\pm 0.15$ eV. 
	Note the difference in vertical scales for some quantities. 
	The transitions are ordered as given in Table S-1 in the SI.}
  \label{Fig-3}
\end{figure*}

We obtained 0-0 energies for 31 transitions at the CC3/\emph{aug}-cc-pVTZ//CC3/\emph{def2}-TZVPP level  (see the SI for details about states, compounds and total energies). Using these values as references, 
we can estimate the errors made while selecting a lighter level of theory for the geometry optimizations, while conserving CC3 for the transition energies. The results are displayed in Figure \ref{Fig-3} and a statistical analysis is provided in Table \ref{Table-1}.

%
% Table-1
% 
\begin{table*}
\caption{Statistical analysis of the impact of the selected method for the geometry optimization on the computed CC3/\emph{aug}-cc-pVTZ transition energies
for 31 representative organic compounds. 
CC3 structures are systematically used as references. 
MSE, MAE, and RMS are given in eV and they correspond to the mean signed, mean absolute and root mean square errors, respectively. 
$\ChemAcc$ and $\AccErr$ are the percentage of cases reaching ``chemical accuracy'' (absolute error $<0.043$ eV) and ``acceptable error'' (absolute error $< 0.150$ eV), respectively.}
\label{Table-1}
\begin{ruledtabular}
\begin{tabular}{llddddd}
Geometry		& Property				& \mcc{MSE}	& \mcc{MAE}	& \mcc{RMS}	& \mcc{$\ChemAcc$} &  \mcc{$\AccErr$} \\
\hline
CCSDR(3)	& $\Ea$							& 0.008	&0.008	&0.011	&100	&100	\\
			& $\Ef$							& 0.024	&0.033	&0.043	&65	&100	\\
			& $(\Ea + \Ef)/2$					& 0.016	&0.020	&0.026	&87	&100	\\
			& $(\EreorgGS - \EreorgES)/2$			&-0.016	&0.020	&0.026	&87	&100	\\
			& $\Eadia$						&0.000	&0.001	&0.001	&100	&100 	\\
\\
CCSD		& $\Ea$							& 0.070	&0.073	&0.086	&32	&90	 \\
			& $\Ef$							& 0.166	&0.173	&0.198	&10		&48	\\
			& $(\Ea + \Ef)/2$					&0.118	&0.123	&0.138	&10		&65	 \\
			& $(\EreorgGS - \EreorgES)/2$			&-0.111	&0.116	&0.130	&10		&68	 \\
			& $\Eadia$						&0.007	&0.007	&0.010	&100	&100 	\\
\\
CC2			& $\Ea$							& -0.074	&0.082	&0.114	&42	&81	\\
			& $\Ef$							&-0.107	&0.130	&0.169	&23	&65	 \\
			& $(\Ea + \Ef)/2$					& -0.090	&0.106	&0.136	&29	&68	\\
			& $(\EreorgGS - \EreorgES)/2$			&0.102	&0.117	&0.149	&26	&65	\\
			& $\Eadia$						&0.011	&0.012	&0.017	&97	&100 	\\
\\
ADC(2)		& $\Ea$							&-0.023	&0.046	&0.064	&58	&94 	\\
			& $\Ef$							&-0.053	&0.122	&0.159	&26	&71 	\\
			& $(\Ea + \Ef)/2$					&-0.038	&0.075	&0.097	&36	&84	 \\
			& $(\EreorgGS - \EreorgES)/2$			&0.060	&0.093	&0.120	&36	&77	 \\
			& $\Eadia$						&0.022	&0.024	&0.037	&74	&100	 \\
\end{tabular}
\end{ruledtabular}
\end{table*}

The bar chart of the error patterns obtained for the vertical absorption and fluorescence are displayed in the top two rows of Figure \ref{Fig-3}.   Although the set of compounds considered here is significantly larger
than the previously studied one, \cite{Jac18a} the major trends and conclusions pertain. For a given method, the errors tend to be significantly larger for $\Ef$ than for  $\Ea$. This unsurprising observation is
due to the higher methodological sensitivity of the ES  geometries compared to their GS counterparts.\cite{Bud17} This is particularly striking with CCSDR(3) that systematically delivers 
chemically-accurate $\Ea$ as compared to CC3, but attain this goal in ``only'' 64.5\%\ of the cases for $\Ef$.  For a given molecule, one notices a general (but not systematic) correlation between the errors made 
for the two kinds of vertical transition energies: if a molecular GS structure is sensitive to the selected method, the same will hold for its ES geometry. Turning now to the comparison
of the four methods [CCSDR(3), CCSD, CC2 and ADC(2)], it is obvious that CCSD geometries yield almost systematically too large absorption and fluorescence energies, with respective MSE of $0.070$ and $0.166$ eV. Qualitatively, 
these positive MSE confirm that CCSD provides a overlocalized picture of the system (in other words, too close from the HF picture), which is consistent with previous works. \cite{Sch08,Bud17,Jac18a}  
With CC2 geometries, the errors go in the opposite direction but their magnitudes are similar to the one obtained with CCSD (MSE of $-0.074$ and $-0.107$ eV for $\Ea$ and $\Ef$, respectively). In other words, 
CC2 yields a too delocalized picture for the geometries, \cite{Bud17,Jac18a} that is, a description with the same error sign as a LDA or GGA functional in the DFT framework. The introduction of perturbative 
triples allows to correct most of the CCSD error, consistently with previous works.\cite{Sau09,Jac18a} Finally, ADC(2) geometries give a more erratic error pattern, but provides a MAE slightly smaller than with CC2
structures for $\Ea$, consistent with the well-known quality of MP2 GS geometries.

Let us now turn our attention to $\Eadia$ and its two contributions as given in Eq. \eqref{eq:2} (bottom three rows of Figure \ref{Fig-3}). It is certainly unsurprising that the error patterns obtained for the average of the absorption and emission
energies (middle row of Figure \ref{Fig-3}) show the same trends as the one described above for absorption and fluorescence, i.e., a limited overestimation with CCSDR(3), a strong 
overshooting with CCSD, a strong underestimation with CC2, and a less clear pattern with ADC(2).  For $(\Ea + \Ef)/2$, chemical accuracy (as compared to CC3 structures) is reached
in 87\%\ of the cases with CCSDR(3), but only for 10\%, 29\%, and 35\%\ of the transitions with CCSD, CC2 and ADC(2), respectively (Table \ref{Table-1}). With the three latter 
methods, there is also a significant share of the cases with errors exceeding 0.15 eV, a large discrepancy given that it solely originates from the structures. Undoubtedly, it is 
striking that these large deviations are nearly exactly compensated by the errors made for the average difference of reorganization energies (penultimate row of Figure  \ref{Fig-3}). Indeed, the third and fourth rows of Figure \ref{Fig-3} are almost perfect mirror image of each other: while CCSD provides too large transition energies, its overlocalized description induces a significant underestimation of the reorganization energies. The same phenomenon 
holds for the other methods, and what was noticed for formaldehyde in the previous Section is therefore a very solid trend. Overall, very small errors are obtained for the adiabatic energies, with trifling MAEs
of $0.001$, $0.007$ and $0.012$ eV with CCSDR(3), CCSD, and CC2 geometries, respectively. Particularly astonishing is the success of CCSD for $\Eadia$ (bottom row of Figure  \ref{Fig-3}): chemical accuracy is systematically reached,
although this is almost never the case for the two components of Eq.~\eqref{eq:2}.  In addition, the error magnitude in $\Eadia$ now follows the expected accuracy ladder: the more refined 
the method, the smaller the average error, which was not necessarily the case for the other properties.

\subsection{Comparisons with experiment for singlet states}

Having performed a theory \emph{versus} theory analysis, let us now turn to comparisons with experimental $\EOO$ values.  We have build a statistically significant list of states for which we have determined CC3 transition energies 
on various geometries. The set encompasses 51 singlet-singlet transitions, extending therefore our previous set and including more challenging cases. \cite{Loo18b}  The full list of data, including experimental references and 
symmetries of all states is available in the SI.  We note that, in a few cases, the ZPVE could not be determined with all levels of theory, typically because some geometry optimizations in a given symmetry failed to converge (see pp {S12--S19} in 
the SI for details).  These cases are not statistically relevant for our purposes.

Before getting to the statistics, let us briefly discuss one interesting example, the first electronic transition in pyrazine. With ADC(2) and CC2, the geometry of the lowest ES belongs to the $C_i$ point group, which is consistent
with a previous investigation performed with these theories. \cite{Win13} The deformation compared to the $D_{2h}$ symmetry is significant with non-equal CN bond lengths attaining 1.288 and 1.368 \AA\ with ADC(2) and 1.294 and
1.373 \AA\ with CC2.  B3LYP/6-31+G(d) foresees a $C_{2h}$ point group, with also strongly divergent CN distances (1.300 and 1.373 \AA).  In contrast, CCSD provides a $D_{2h}$ ES geometry --- all CN bonds being 1.338 \AA\ long ---
which is consistent with higher-level of theories, as the CCSDR(3) optimization starting from the B3LYP minimum leads back to a $D_{2h}$ structure with CN distances of 1.346 \AA.  Despite these rather divergent
descriptions of the ES geometries, the CC3 0-0 energies are always accurate with deviations of 0.031 eV, 0.023 eV, 0.036 eV and 0.038 eV compared to experiment when using the ADC(2), CC2, CCSD 
and CCSDR(3) minimum, respectively.   In other words, a chemically-accurate $\EOO$ can be obtained even with an incorrect point group symmetry for the ES geometry.  As discussed in our recent work, \cite{Loo18b} changing the level of theory for computing  
the $\Eadia$ transition is much more deleterious, e.g., the CCSD//CCSD $\EOO$ is 0.157 eV off the experimental value.
 
%
% Table-2
% 
\begin{table*}
\caption{Statistical data obtained by comparing experimental and theoretical $\EOO$ values for singlet-singlet transitions. 
The transition energies are systematically computed at the CC3/\emph{aug}-cc-pVTZ level using different geometries and ZPVE corrections. 
The latter term computed at the \emph{def2}-TZVPP basis set for all methods, except for B3LYP for which the more compact 6-31+G(d) basis set was applied. 
``Count'' refers to the number of transitions in each group.
$\ChemAcc$ and $\AccErr$ are the percentage of cases reaching ``chemical accuracy'' (absolute error $<0.043$ eV) and ``acceptable error'' (absolute error $< 0.150$ eV), respectively.}
\label{Table-2}
\begin{ruledtabular}
\begin{tabular}{lldddddd}
Geometry		& ZPVE				& \mcc{Count}		& \mcc{MSE}	& \mcc{MAE}	& \mcc{RMS}	& $\ChemAcc$ &  $\AccErr$ \\
\hline
CC3			& CCSD				&	33		& -0.014	& 0.028	& 0.038	&76	& 100	\\
			& B3LYP				&	33		& -0.014	& 0.025	& 0.038	&82	& 100	\\
\\	
CCSDR(3)	& CCSD				&	48		& -0.010	& 0.027	& 0.036	&79	& 100 	\\
			& B3LYP				&	48		& -0.009	& 0.024	& 0.035	&83	& 100	\\
\\
CCSD		& CCSD				&	51		& -0.002	& 0.025	& 0.033	& 84	& 100	\\	
			& CC2				&	51		& -0.004	& 0.026	& 0.034	& 78	& 100	\\
			& ADC(2)				&	51		& -0.007	& 0.029	& 0.037	& 77	& 100	\\
			& B3LYP				&	51		& -0.002	& 0.024	& 0.032	& 86	& 100	\\
\\
CC2			& CC2				&	51		&0.004	& 0.032	& 0.045	& 77	& 98	\\
			& ADC(2)				&	51		&0.002	& 0.033	& 0.049	& 75	& 98	\\
			& B3LYP				&	51		&0.006	&0.030	&0.042	& 78	&98	\\
\\
ADC(2)		& ADC(2)				&	51		& 0.022	& 0.048	& 0.064	& 61	& 98	\\
			& B3LYP				&	51		& 0.027	& 0.048	& 0.066	& 65	& 96	\\
\end{tabular}
\end{ruledtabular}
\end{table*}

Table \ref{Table-2} reports statistical results obtained for the set of considered compounds, combining various levels of theory for the geometry optimization and the calculation of the $\Ezpve$
correction term.  With the most refined level of theory, that is, selecting CC3 geometries and CCSD ZPVE, one obtains a MAE of 0.028 eV and a RMS of 0.038 eV, both below the chemical accuracy threshold.  
This approach also delivers errors smaller than 0.15 eV for all compounds, and reaches chemical accuracy in 76\%\ of the cases, a success slightly less impressive than in our previous contribution, \cite{Loo18a}  but the
present set contains many transitions of $n \rightarrow \pi^\star$ character that induce large geometrical reorganization often harder to capture with theory, as well as several, ``exotic'' species, such as \ce{HPO}, \ce{HPS}, and 
\ce{H2C=Si}. For instance, the CC3//CC3 $\EOO$ comes with a significant error of $-0.102$ eV compared to experiment for \ce{HPS}, but even the CAS/MRCI+D/aug-cc-pV(5+d)Z method suffers from a (relatively) large 
error (0.073 eV) for this compound. \cite{Gri13b} As one can see in Table \ref{Table-2}, using B3LYP ${\Ezpve}$ corrections instead of the CCSD ones essentially induces no change. This conclusion is well 
in line with the analysis of Winter and coworkers who demonstrated the small impact of the level of theory used for determining the vibrational correction. \cite{Win13} When using CCSDR(3) or CCSD geometries, one also 
notices very small changes of the statistical values, the CCSD geometries delivering actually slightly more accurate results than the CC3 one, with a chemical accuracy rate reaching an astonishing 86\%\ with the B3LYP ZPVE,
close to the one we reported earlier. \cite{Loo18a} As stated above, \cite{Bud17} the CCSD ES geometries are certainly not very accurate, so this illustrates that a very nice match between experimental and theoretical $\EOO$ 
values can be reached even with not-so-accurate structures.  When one climbs down the accuracy scale to CC2 geometries, the results  remain very acceptable with a RMS close to chemical accuracy, whereas using ADC(2) for 
the geometry optimizations further degrades the results.

\subsection{Triplet excited states}

For the triplet transitions, several methodological choices can be made. First, given the above results, we consider CCSD geometries only and computed the $\Ezpve$ corrections at this level as well as with B3LYP.  In addition, as the collected experimental data 
nearly systematically focus on the lowest triplet state of a given symmetry, we optimize the corresponding geometries at the U-CCSD/\emph{def2}-TZVPP level, which offers a more computationally appealing approach than EOM-CCSD.  
For the adiabatic energies, we test both the LR-CC3/\emph{aug}-cc-pVTZ approach as implemented in Dalton,  \cite{dalton} and the restricted open-shell ``ground state'' implementation available in Psi4. \cite{Psi4}  In this latter approach the initial ROHF 
reference orbitals are transformed into semi-canonical UHF orbitals during the calculation. Twenty-two singlet-triplet transitions have been considered in total. As above, details regarding total and transition energies, symmetries, and experimental references 
can be found in the SI.

First, as can be deduced from Tables S-21 and S-22, the differences obtained by using LR-CC3 or RO-CC3 energies are generally insignificant. Indeed, for the full set of compounds, the MSE (MAE) between the two
series of adiabatic energies is as small as 0.018 eV (0.028 eV). There is, however, two cases in which significant changes are noticed: acetylene ($+0.060$  eV) and cyanogen ($+0.300$ eV). Although there is
no obvious rationale for these larger discrepancies, we note that these two outliers are the only systems of our set for which the lowest singlet-triplet transition as a $\pi \rightarrow \pi^\star$ rather than a $n \rightarrow \pi^\star$ character. Next,
few specific molecules are worth further discussion.  The first is \ce{CHCl}, for which we computed a negative vertical emission energy. This unsettling feature is due to a crossing between the $S_0$ and $T_1$ potential energy surfaces, i.e., for 
the GS $S_0$ geometry, $S_0$ is, of course, the most stable state, whereas for the optimal triplet geometry, the lowest triplet indeed becomes energetically favored compared to the singlet. We note that the computed $\EOO$ is chemically 
accurate for that compound irrespective of the selected protocol. The second molecule is acetylene, for which our theoretical $\EOO$ values (3.752--3.814 eV) deviate substantially from a rather recent experimental value of 3.584 eV, 
\cite{Ahm99} but fit very well previous MR-AQCC values (3.84 eV), \cite{Ven03} or theoretical best estimates (3.81 eV).\cite{She00} Consistently with the detailed analyses carried out by Sherrill, Head-Gordon, Schaefer,
and their coworkers, \cite{Lun93,She00} we therefore conclude that the experimental value is inaccurate and we discard acetylene from our statistics. Third, there is \ce{SO2} for which the errors are abnormally large, e.g., $-0.218$ eV
with LR-CC3. Given previous studies on this molecule showing unusually large basis set effects, e.g., see Ref. \citenum{Fel99} and references therein, we performed LR-CC3/d-\emph{aug}-cc-pVQZ calculations, which allowed halving 
the error ($-0.109$ eV). For the sake of consistency, we have nevertheless kept the original \emph{aug}-cc-pVTZ result in our statistics.

The statistical data obtained for the singlet-triplet transitions are given in Table \ref{Table-3}. As can be seen, the four tested protocols provide similar deviations and there is no advantage nor disadvantage of using RO-CC3 instead of
LR-CC3, whereas the improvement brought by using CCSD ZPVE corrections instead of their B3LYP counterparts is very small, which probably does not justify the associated increase in computational cost.  The MSEs are negative and 
there is indeed a clear tendency to (slightly) underestimate the experimental values for singlet-triplet transitions. The MAEs are larger than for the singlet-singlet transitions, but this is mainly due to SO$_2$. Indeed, removing it from the set
would decrease the LR-CC3 MAE from 0.039 to 0.030 eV, close to the 0.025 eV value obtained in the previous Section with an equivalent approach. Chemical accuracy is reached in 76\%\ of the cases with both LR-CC3//U-CCSD
and RO-CC3//U-CCSD approaches, which is certainly a very pleasant outcome for spin-flip transitions.
%
% Table-3
% 
\begin{table*}
\caption{Statistical data obtained by comparing experimental and theoretical $\EOO$ values for singlet-triplet transitions. 
The transition energies are systematically computed at the CC3/\emph{aug}-cc-pVTZ level using different protocols and ZPVE corrections. 
The latter term was computed at the \emph{def2}-TZVPP basis set (CCSD) or 6-31+G(d) (B3LYP) level. 
``Count'' refers to the number of transitions in each group.
$\ChemAcc$ and $\AccErr$ are the percentage of cases reaching ``chemical accuracy'' (absolute error $<0.043$ eV) and ``acceptable error'' (absolute error $< 0.150$ eV), respectively.}
\label{Table-3}
\begin{ruledtabular}
\begin{tabular}{llldddddd}
Geometry	& $\Eadia$	& ZPVE			&	\mcc{Count}	& \mcc{MSE}	& \mcc{MAE}	& \mcc{RMS}	& \mcc{$\ChemAcc$} &  \mcc{$\AccErr$} \\
\hline
U-CCSD	&	LR-CC3	& U-CCSD			&21		&-0.032	&0.039	&0.059	&76	&95	\\
		&	LR-CC3	& U-B3LYP			&21		&-0.040	&0.041	&0.062	&76	&95	\\
		&	RO-CC3	& U-CCSD			&21		&-0.016	&0.046	&0.074	&76	&91	\\
		&	RO-CC3	& U-B3LYP			&21		&-0.024	&0.046	&0.074	&76	&91	\\
\end{tabular}
\end{ruledtabular}
\end{table*}

\subsection{Radicals}

Let us now turn towards the calculation of $E^{0-0}$ for radical species. Open-shell molecules are more challenging for theoretical methods than their closed-shell counterparts, and this certainly holds for ES properties. 
In particular, DFT and TD-DFT are known to be less effective for radicals, or at the very least, the ``optimal'' functional for  $E^{0-0}$ are different for open- and closed-shell molecules.\cite{Die04b} We nevertheless 
computed $\Ezpve$ with B3LYP/6-31+G(d) when technically feasible (a few cases did not converge, see the SI).  As for triplets, we use the unrestricted formalism during the geometry optimization, 
i.e.,  the GS and ES structures are obtained at the U-CCSD/\emph{def2}-TZVPP and  U-EOM-CCSD/\emph{def2}-TZVPP levels, respectively. For the calculations of the CC3 transition energies, we apply the restricted open-shell
protocol implemented in Psi4. \cite{Psi4} All the total and transition energies as well as geometries are available in the SI.

In contrast to the previous cases, the term ${\Ezpve}$ is not systematically negative for radicals, it can be very close to zero (\ce{F2BO}) or even significantly positive (\ce{NO3}, vinyl, \ldots). We obtain a very reasonable 
agreement between theory and experiment for most radicals (\emph{vide infra}), although for two cases, \ce{CNO} and \ce{FS2}, the deviations exceed 0.150 eV. For the former compound,  a previous MRCI+Q/cc-pVQZ investigation 
reported a smaller error for $\EOO$ with respect to experiment. \cite{Leo08}. We therefore computed CC3/\emph{aug}-cc-pVQZ estimates, but the changes were trifling compared to the triple-$\zeta$ basis set, 
therefore hinting that multi-reference effects are probably significant for \ce{CNO}. Interestingly, for the \ce{NCO} isomer, our approach is chemically accurate. For \ce{FS2}, we did not found any high-level multi-reference results in 
the literature to compare with, so that the origin of the theory/experiment discrepancy could not be clarified.  We suspect here large basis set effects as those noted for SO$_2$, though it was technically beyond reach to
ascertain this guess.

As expected for these more challenging compounds, the average deviations are larger than for the closed-shell species (see Table \ref{Table-4}). Nevertheless the MAE remains close to chemical accuracy and the theoretical
prediction matches that target accuracy in two third of the cases, with only two compounds out of twenty for which the error exceeds 0.150 eV, the typical average error of standard ES approaches (see the Introduction).

%
% Table-4
% 
\begin{table*}
\caption{Statistical data obtained by comparing experimental and theoretical $\EOO$ values in radials. 
The transition energies are systematically computed at the LR-RO-CC3/\emph{aug}-cc-pVTZ level using two different ZPVE corrections. 
The latter term was computed at the \emph{def2}-TZVPP basis set (CCSD) or 6-31+G(d) (B3LYP) level.
``Count'' refers to the number of transitions in each group.
$\ChemAcc$ and $\AccErr$ are the percentage of cases reaching ``chemical accuracy'' (absolute error $<0.043$ eV) and ``acceptable error'' (absolute error $< 0.150$ eV), respectively.}
\label{Table-4}
\begin{ruledtabular}
\begin{tabular}{lldddddd}
Geometry	& ZPVE			& \mcc{Count}		& \mcc{MSE}	& \mcc{MAE}	& \mcc{RMS}	& $\ChemAcc$ &  $\AccErr$ \\
\hline
U-CCSD	& U-CCSD			&20		& 0.023	&0.043	&0.063	&65	&90	\\
		& U-B3LYP			&20		& 0.018	&0.041	&0.062	&70	&90	\\
\end{tabular}
\end{ruledtabular}
\end{table*}

\section{Conclusions}

\begin{figure}
  \includegraphics[width=\linewidth]{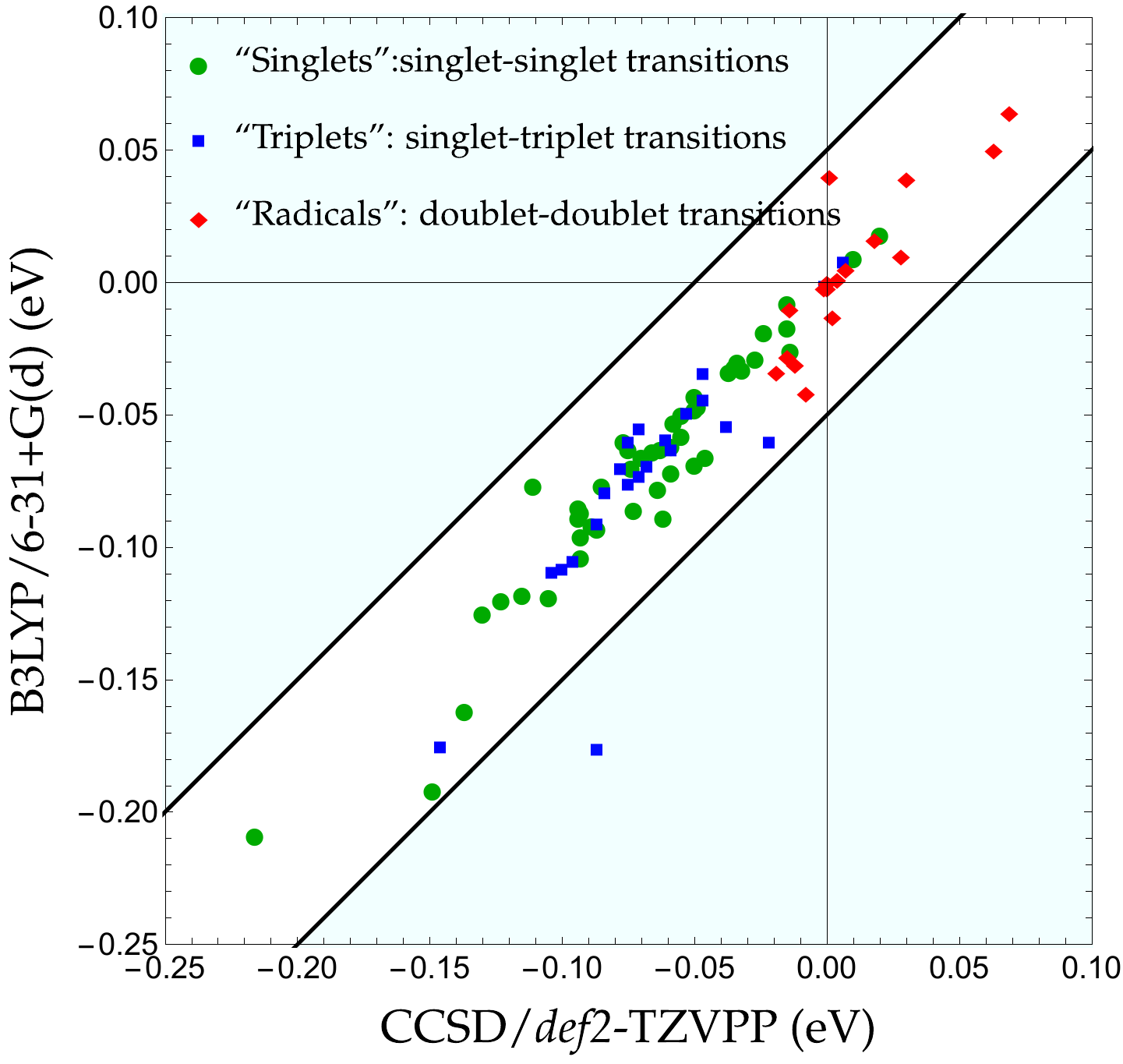}
  \caption{Comparisons of $\Ezpve$ (eV) computed with B3LYP/6-31+G(d) and CCSD/\emph{def2}-TZVPP for the three subsets of compounds: ``singlets'' (green dots), ``triplets'' (blue squares) and ``radicals'' (red diamonds).
  The white zone delimited by the two black lines indicates an absolute deviation smaller than $0.05$ eV.}
  \label{Fig-4}
\end{figure}

\begin{figure}
  \includegraphics[width=\linewidth]{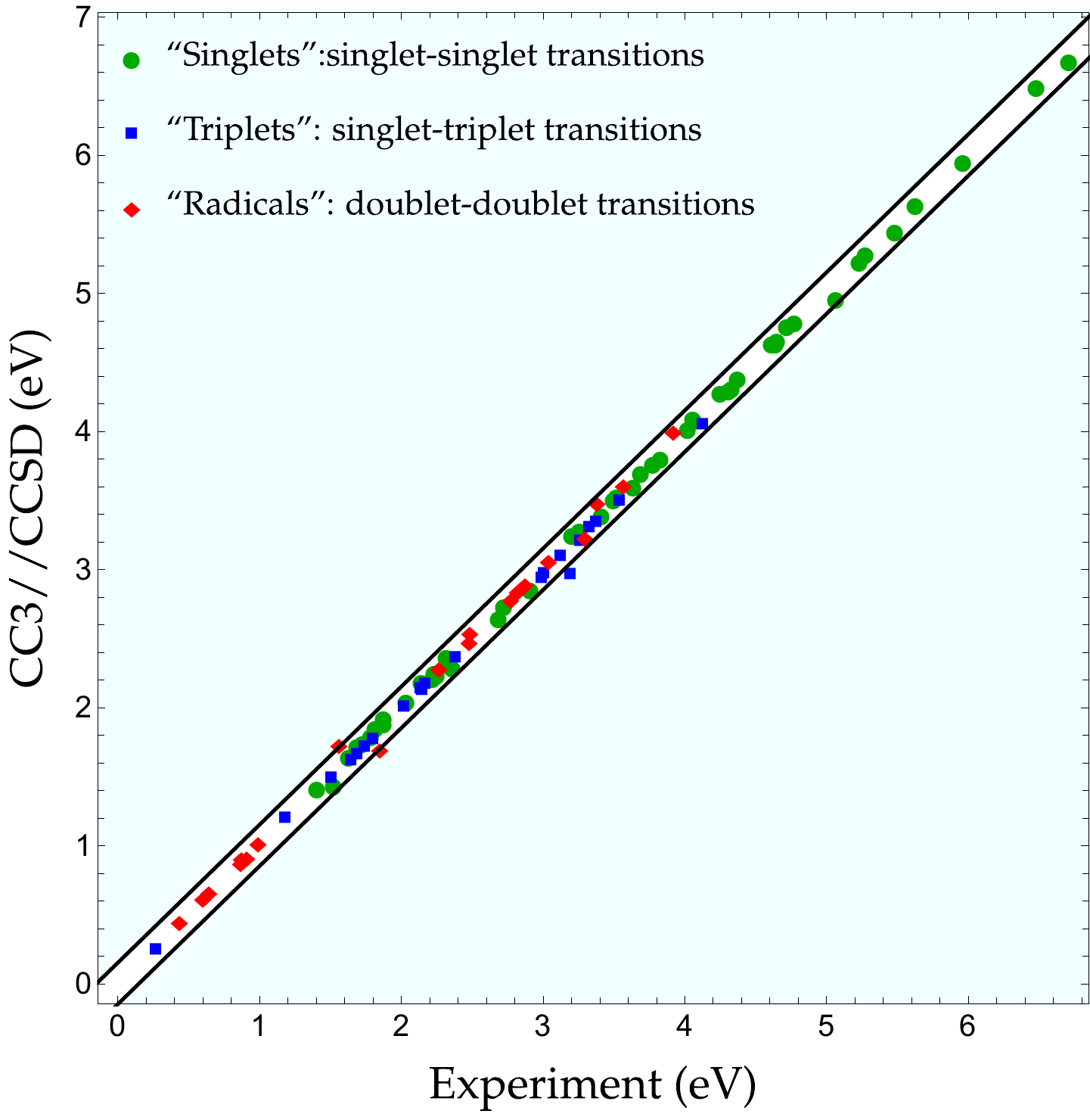}
  \caption{Comparison between experimental and theoretical 0-0 energies for the three subsets of compounds: ``singlets'' (green dots), ``triplets'' (blue squares) and ``radicals'' (red diamonds).
  The white zone delimited by the two black lines indicates an absolute deviation smaller than $0.15$ eV.}
  \label{Fig-5}
\end{figure}

We have computed 0-0 energies in nearly 100 compounds using a panel of increasingly accurate wavefunction approaches. In contrast to previous benchmark studies devoted to $\EOO$, our focus was set on the level of theory 
used to determine the ground and excited state geometries as well as zero-point corrections, rather than the transition energies. For the latter, we systematically applied the CC3/\emph{aug}-cc-pVTZ level, correlating all electrons 
to provide an accurate and uniform description.  For the ZPVE correction term, ${\Ezpve}$, we found a very good agreement between the values obtained at various levels of theories, so that one can safely use
a computationally lighter approach for such a quantity. For instance, for the 86 cases in which both B3LYP/6-31+G(d) and CCSD/\emph{def2}-TZVPP  ${\Ezpve}$ values could be determined, we found a mean absolute deviation
between the two as small as 0.01 eV. The correlation between the two sets of data is also obvious, with only one case for which the deviation exceeds 0.05 eV (see Figure \ref{Fig-4}). The influence of the geometry is rather small as well. Using CC3,
CCSDR(3) or CCSD geometries essentially yield the same statistical deviations, a very small drop in accuracy being noticed for the CC2 structures and a more substantial one for the ADC(2) geometries.
By comparing the experimental and theoretical 0-0 energies obtained by combining i) CC3 adiabatic energies, ii) CCSD geometries, and iii) B3LYP ZPVE corrections, we could reach chemical accuracy in
80.4\%\ of the cases with a trifling MSE of $-0.006$ eV and a MAE of $0.031$ eV. As highlighted in Figure \ref{Fig-5}, it is probably even more striking that this success is obtained for $\EOO$ covering a wide range of
energies. Even though our protocol is limited to single-reference methods and triple-$\zeta$ basis set, only three cases (out of 92) were found having a theory-experiment discrepancy exceeding $0.15$ eV 
(the lowest triplet of \ce{SO2}, and the doublet-doublet transitions in \ce{CNO} and \ce{FS2}). A key observation of the present study is that this unexpected success is a direct consequence of a strong and systematic error 
cancelation between the vertical transition energies and the CCSD geometry reorganization energies. It was indeed found previously that the CCSD geometries, and hence the vertical absorption and emission energies computed on these 
structures, \cite{Jac18a} are rather poor with significant deviations compared to the reference values for polar bonds. \cite{Bud17} Therefore, while one can rely on this solid error compensation phenomenon to determine chemically-accurate 
$\EOO$ with ``cheap'' geometries, the drawback is that a close match between experiment and theory is no proof of the geometry accuracy.

To sum up, while we confirm our previous conclusions that chemical accuracy can be reached using CC3 to compute $\Eadia$, \cite{Loo18b} we additionally demonstrate that this does not necessarily requires very accurate geometries 
as generally thought. 

\begin{acknowledgements}
D.J.~acknowledges the \emph{R\'egion des Pays de la Loire} for financial support. This research used resources of i) the GENCI-CINES/IDRIS (Grant 2016-08s015); 
ii) CCIPL (\emph{Centre de Calcul Intensif des Pays de Loire}); and iii) the Troy cluster installed in Nantes.
\end{acknowledgements}

\section*{Supporting Information}
List of total and transition energies for all compounds, list of experimental values and references, CCSD ground and excited state geometries.

\end{document}